\documentclass[a4paper,11pt]{article}
\pdfoutput=1
\usepackage{amsmath}
\usepackage{amssymb}
\usepackage{color}
\usepackage{cite}
\usepackage{graphicx}
\usepackage{leftidx}
\usepackage{subfig}
\usepackage{hyperref}

\numberwithin{equation}{section}

\def\beq{\begin{equation}}
\def\eeq{\end{equation}}

\def\l{\lambda}

\def\o{\omega}
\def\O{\Omega}

\def\d{\mathrm{d}}

\newcommand{\ben}{\begin{enumerate}}
\newcommand{\een}{\end{enumerate}}
\newcommand{\be}{\begin{equation}}
\newcommand{\ee}{\end{equation}}
\newcommand{\mc}{\mathcal}
\newcommand{\mr}{\mathrm}

\newcommand{\de}{\delta}
\newcommand{\del}{\partial}
\newcommand{\w}{\wedge}
\newcommand{\da}{\dagger}

\usepackage{tikz}
\usetikzlibrary{positioning}
\usetikzlibrary{intersections}
\usetikzlibrary{fadings} 
\usetikzlibrary{arrows.meta} 
\usetikzlibrary{arrows}

\tikzfading[name=fade out,
inner color=transparent!0,
outer color=transparent!100]

\definecolor{cherryblossompink}{rgb}{1.0, 0.72, 0.77}
\definecolor{lightblue}{rgb}{0.68, 0.85, 0.9}

\usetikzlibrary{decorations.pathmorphing}
\usetikzlibrary{decorations.pathreplacing,decorations.markings}

\usetikzlibrary{backgrounds,automata}

\begin{document}

\title{\textbf{Canonical quantization for effective theories with perturbations altering degrees of freedom: a covariant phase space approach}}
\author{
Jie-qiang Wu$^{1,2}$\footnote{jieqiangwu@itp.ac.cn}\, and
Jinan Zhao$^{1,2}$\footnote{jinanzhao@itp.ac.cn}
}
\date{\today}

\maketitle

\begin{center}
{\it
$^{1}$Institute of Theoretical Physics, Chinese Academy of Sciences, Beijing 100190, China \\
$^{2}$School of Physical Sciences, University of Chinese Academy of Sciences, \\ Beijing 100049, China
}
\vspace{10mm}
\end{center}

\begin{abstract}

The standard approach to canonical quantization encounters difficulties in dealing with perturbations that alter the kinetic structure of unperturbed theories. We show that the covariant phase space formalism provides a natural and technically efficient way to circumvent this obstruction. We illustrate the method with an exactly solvable model: a two-dimensional non-relativistic charged particle moving in a magnetic field and a harmonic confining potential, with its kinetic energy viewed as a perturbation. We quantize this model with covariant phase space formalism by constructing the solution perturbatively. We then calculate the energy spectrum and the unequal-time commutators of this model, and obtain the results that agree with the expansion of the exact theory. The procedure developed here is intended to serve as a systematic framework for the canonical quantization of more complex effective theories with higher-derivative or velocity-dependent perturbations.

\end{abstract}

\baselineskip 18pt
\thispagestyle{empty}

\newpage

\tableofcontents
\newpage

\section{Introduction}\label{Sec1}

Canonical quantization for theories 
in which perturbations alter degrees of freedom
presents a technical challenge.
For example, in an unperturbed constraint theory, one identifies canonical momenta and constraints of the system, and quantizes it using the standard Dirac-Bergmann algorithm\cite{Bergmann:1949zz,Dirac:1950pj,Dirac:1951zz,Anderson:1951ta,Bergmann:1954tc,Dirac1964lectures,Brown:2022gha,Brown:2022iez}. 
However, when a velocity-dependent perturbation is added to the Lagrangian, the kinetic structure of the theory can be modified, and what was a constrained system in the unperturbed theory becomes an unconstrained system in the perturbed theory. These two theories live on phase spaces of different dimensions, and their canonical commutation relations are distinct. Consequently, in the standard approach to canonical quantization, trying to quantize such systems in a perturbative way runs into fundamental difficulties.

Building on earlier ideas from \cite{Witten:1986qs,Zuckerman:1986vzu,Crnkovic:1987,Crnkovic:1987tz}, the covariant phase space (CPS) formalism developed by Iyer, Lee, Wald and Zoupas\cite{Lee:1990nz,Wald:1993nt,Iyer:1994ys,Iyer:1995kg,Wald:1999wa} offers a conceptually different path to canonical quantization. Instead of starting from a $3+1$ decomposition of spacetime, computing conjugate momenta and constructing constraints, this formalism establishes the phase space directly from solutions to equations of motion (EOMs), and equips it with a closed symplectic form. The symplectic form in CPS automatically captures the canonical structure of the system, without reference to constraints in the theory
\footnote{We refer readers to the appendix A in the forthcoming work\cite{Gao:2026:AdS3Proca} for the equivalence between the CPS and the Dirac-Bergmann formalism in the case that there are no gauge redundancies.}.  
We then perform quantization by promoting the inverse of the symplectic form to a commutator, and obtain the quantum theory whose commutation relations faithfully inherit the canonical structure of the classical theory.

The purpose of this paper is to demonstrate how CPS realizes canonical quantization for theories in which perturbations alter degrees of freedom.
We illustrate the method with an exactly solvable model whose unperturbed limit possesses second-class constraints, while constraints are absent in the full theory.
We adopt the viewpoint of effective field theory: perturbations in the Lagrangian are regarded as low-energy remnants of a more fundamental theory, and the effective theory is only valid up to a finite cutoff scale\cite{Cheng:2001du,Jaen:1986iz}. In this framework, the perturbative expansion is organized around the low-energy degrees of freedom, and the fast modes that lie beyond the cutoff are expected to decouple.  When gauge redundancies are absent, given a Lagrangian with a perturbation that modifies the kinetic structure, the perturbative quantization procedure developed in this paper can be summarized as follows:
\begin{enumerate}
    \item We take a variation of the Lagrangian, read off the EOMs as well as the symplectic potential $\Theta$, and construct the symplectic form $\O$ together with the Hamiltonian $H$ defined as the Noether charge associated with time translation.
    \item We write the solution to EOMs as a perturbative expansion, and solve corrections to the unperturbed solution order by order by expanding EOMs in terms of the perturbation parameter.
    \item We substitute the perturbative solution into the symplectic form, and quantize the system by promoting the inverse of the symplectic form to a commutator.
    \item We calculate physical quantities, such as energy spectrum and correlators, based on the commutation relations obtained above.
\end{enumerate}

The feasibility of perturbative canonical quantization with CPS was first demonstrated by Basu in a preceding work\cite{Basu:2004qy}, which analyzed a quartic oscillator with a standard, non-singular kinetic term. In that model, the perturbation does not alter the dimension of the phase space nor the structure of the canonical commutation relations. The main technical challenge lies in the deformation of the Poisson brackets, and the construction of dressed canonical variables that restore the standard canonical commutation relations order by order. In this paper, we address a more challenging scenario in which the perturbation modifies the kinetic structure of the Lagrangian, and changes the constraints as well as the dimension of the phase space. This is a qualitatively different obstruction, and one cannot treat the perturbation as a mere deformation of a fixed algebraic framework.

The rest of this paper is organized as follows: 
In Sec.~\ref{Sec2} we present the Lagrangian of this model and identify $\l$ as the perturbation parameter. 
In Sec.~\ref{Sec3} we apply the standard canonical quantization procedure to the model. We first treat the $\l = 0$ case using the Dirac-Bergmann algorithm. We then quantize the $\l > 0$ case as an exact theory. We show that, although the exact theory correctly reduces to the $\l = 0$ theory in the limit $\l \to 0$, treating $\l$-term as a perturbation within the standard framework of canonical quantization is not feasible. 
In Sec.~\ref{Sec4} we reformulate the quantization with CPS. We construct the symplectic form directly on the solution space, solve the EOMs order by order in $\l$ using perturbation theory, and show that quantization via the inverse of the symplectic form yields the quantum theory that agrees with the expansion of the exact theory. 
In Sec.~\ref{Sec5} we summarize our results and discuss their implications.
Technical details of derivations omitted in the main text are collected in Appendices~\ref{appA}, \ref{appB} and \ref{appC}.

In this paper we work in the Heisenberg picture. We use natural units with $\hbar = c = 1$. We add dots above functions to represent time derivatives, i.e., 
$\dot{x}(t) = \frac{\d x(t)}{\d t}$, 
$\ddot{x}(t) = \frac{\d^2 x(t)}{\d t^2}$. We denote the exterior derivative on the configuration space (and also its pullback to the phase space) by $\de$ to distinguish it from the exterior derivative $\d$ on spacetime.

%%%%%%%%%%%%%%%%%%%%%%%%%%%%%%%%%%%%%%%%%%%%%%%%%%%%%%

\section{Lagrangian}\label{Sec2}

The complete Lagrangian of this model is
\begin{equation}\label{Lag}
    \mathcal{L}(x,y,\dot{x}, \dot{y}) = 
     - \frac{\o}{2} (x^2 + y^2)
     - \frac{1}{2} (x \dot{y} - y \dot{x})
     + \frac{\l}{2} (\dot{x}^2 + \dot{y}^2),
\end{equation}
where $\o> 0$ and $\l \ge 0$. 
In the canonical quantization procedure of this system, we would like to treat the $\l$-term in the above Lagrangian as a perturbation.

This model is chosen because it exhibits essential structural features that obstruct the standard perturbative treatment: a singular $\l=0$ limit with second-class constraints, and a velocity-squared perturbation that modifies the constraint structure. At the same time, this model remains sufficiently simple that both the exact solution and the perturbative solution can be obtained analytically. In the following discussion, the exact theory for $\l>0$ will serve as a benchmark against which the perturbative results obtained with CPS should be tested.

This model admits a natural physical interpretation that clarifies in which case the $\l$-term can be regarded as a perturbation. In the symmetric gauge, the Lagrangian of a 2-dimensional non-relativistic charged particle moving in a uniform magnetic field perpendicular to the plane, and confined by a harmonic potential, takes the form
\begin{equation}
    \mathcal{L} =
    -\frac{k}{2}(x^2 + y^2) 
    - \frac{qB}{2}(x\dot{y} - y\dot{x})
    +\frac{m}{2}(\dot{x}^2 + \dot{y}^2),
\end{equation}
where the first term is the harmonic potential, the second term is the magnetic coupling, and the third term is the kinetic energy. In the strong field limit, the kinetic energy becomes a small correction relative to interaction terms
\footnote{The Lagrangian presented here is a formal model designed to illustrate the quantization method. Its physical interpretation as a non-relativistic charged particle is only motivational. In particular, we work in a parameter regime where the characteristic velocity remains sufficiently below the speed of light. And the non-relativistic treatment would not necessarily apply in all parameter regimes without further justification.}. 
Hence, treating $\l$ as a perturbation parameter in the model Eq.~(\ref{Lag}) is physically well-motivated.

%%%%%%%%%%%%%%%%%%%%%%%%%%%%%%%%%%%%%%%%%%%%%%%%%%%%%%

\section{Standard approach to canonical quantization}\label{Sec3}

In this section we apply the standard canonical quantization procedure to this model in two different cases: $\l = 0$ and $\l > 0$. If $\l = 0$, there exist second-class constraints, and we quantize the system via the Dirac–Bergmann algorithm. If $\l > 0$, there are no constraints, and we quantize the system by promoting Poisson brackets to commutators. We argue that in the limit $\l \to 0$, the fast mode decouples from the low-energy physics, and the low-energy sector of the quantum theory for $\l > 0$ reduces to the $\l = 0$ theory. However, because the perturbation modifies the kinetic structure of the system, and consequently alters the constraints, treating the $\lambda$-term as a perturbation within the standard approach to canonical quantization is not feasible.

\subsection{$\l=0$}

\subsubsection{Constraints and Dirac brackets}
\label{Sec3.1.1}

To quantize the system, first we calculate the canonical momenta $\pi_x$ and $\pi_y$
\begin{align}
    \pi_x = \tfrac{\del{\mc{L}}}{\del \dot{x}} 
    &= \tfrac{1}{2} y + \l \dot{x}, 
    \label{canonical momentum x}  \\
    \pi_y = \tfrac{\del{\mc{L}}}{\del{\dot{y}}}
    &= -\tfrac{1}{2} x + \l \dot{y}
    \label{canonical momentum y}.
\end{align}
If $\l=0$, $\dot{x}$ and $\dot{y}$ can not be solved in terms of $x$, $y$, $\pi_x$ and $\pi_y$ from above equations. As a result, there exist two primary constraints
\begin{align}
    \chi_1 &\equiv \pi_x - \tfrac{1}{2} y \approx 0, 
    \label{constraint1} \\
    \chi_2 &\equiv \pi_y + \tfrac{1}{2} x \approx 0,
    \label{constraint2} 
\end{align}
where $\approx$ means that the equality holds only on the constraint surface. Next we calculate the Poisson bracket of these constraints
\begin{equation}
    C_{12} = -C_{21} 
    = [\chi_1, \chi_2]_{\mr{P}} 
    = -1 \ne 0.
\end{equation}
and, of course
\begin{equation}
    C_{11} = C_{22} = 0.
\end{equation}
Therefore, there are no secondary constraints since the matrix $C_{MN}$ is reversible. And the constraints are second-class. The Dirac bracket of two variables $A$ and $B$ is defined as\cite{Dirac1964lectures}
\begin{equation}\label{Dirac bracket}
    [A, B]_{\mathrm{D}} = [A, B]_{\mathrm{P}} 
    - \sum_{N,M} [A, \chi_N]_{\mathrm{P}} 
    (C^{-1})^{NM} [\chi_M, B]_{\mathrm{P}}.
\end{equation}
Evaluating the Dirac brackets gives the following results  (see Appendix~\ref{appA} for detailed computations): 
\begin{align}
[x, y]_{\mathrm{D}} &= 1, & [x, \pi_x]_{\mathrm{D}} &= \tfrac{1}{2}, &
[x, \pi_y]_{\mathrm{D}} &= 0, \notag \\
[y, \pi_x]_{\mathrm{D}} &= 0, & [y, \pi_y]_{\mathrm{D}} &= \tfrac{1}{2}, &
[\pi_x, \pi_y]_{\mathrm{D}} &= \tfrac{1}{4}.
\label{eq:DiracBrackets}
\end{align}

\subsubsection{Quantization}

The EOMs of the system are given by
\begin{align}
    - \dot{y} - \o x &= 0, \label{unperturbed_EOM1}  \\
    \dot{x} - \o y &= 0.  \label{unperturbed_EOM2} 
\end{align}
We eliminate one variable from above equations to obtain
\begin{equation}
    \ddot{x} + \o^2 x = 0.
\end{equation}
And the general solution to this equation can be expanded as
\begin{equation}\label{x0}
    x_0(t) =
    \frac{1}{\sqrt2} 
    ( a e^{-i \o t} + a^{\da} e^{i \o t} ),
\end{equation}
where $a$ and $a^\da$ are complex constants to be determined by initial conditions. We also have
\begin{equation}\label{y0}
    y_0(t) = \o^{-1} \dot{x} 
    = \frac{-i}{\sqrt2} 
    ( a e^{-i \o t} - a^{\da} e^{i \o t} ).
\end{equation}

To quantize the system, the commutation relation of two variables $A, B$ is given by\cite{Dirac1964lectures,Weinberg:1995mt}
\begin{equation}
    [A,B] = i [A,B]_\mr{D}.
\end{equation}
We promote $x_0$ and $y_0$ to operators, and impose the following equal-time commutation relation
\begin{equation}
    [x_0,y_0] = i.
\end{equation}
From the mode expansions Eqs.~(\ref{x0}) and (\ref{y0}) we obtain the commutation relation between operators $a$ and $a^\da$
\begin{equation}\label{a_com}
    [a,a^{\da}] = 1.
\end{equation}
Therefore $a$ and $a^\da$ obey the usual commutation relation of creation and annihilation operators.
The canonical Hamiltonian of this system reads
\begin{equation}
    H_0 
    = \dot{x} \pi_x + \dot{y} \pi_y - \mc{L}
    = \frac{\o}{2} (x^2 + y^2).
\end{equation}
And after quantization, the Hamiltonian can be rewritten in terms of operators $a$ and $a^\da$ as follows
\begin{equation}
    H_0 = \o ( a^\da a + \tfrac{1}{2}).
\end{equation}
We define the number operator $N \equiv a^\da a$, and denote an eigenstate of $N$ by its non-negative integer eigenvalue $n$
\begin{equation}
    N | n \rangle
    = n | n \rangle.
\end{equation}
Thus the energy spectrum of this system is
\begin{equation}
    H_0 | n \rangle
    =  \o ( n + \tfrac{1}{2} ) | n \rangle.
\end{equation}

\subsubsection{Unequal-time commutators}

Next we would like to calculate unequal-time commutators 
$[x_0(t_1), x_0(t_2)]$, 
$[x_0(t_1), y_0(t_2)]$, and 
$[y_0(t_1), y_0(t_2)]$ 
in the quantum theory,
since these objects determine the linear
response of the system to external perturbations\cite{Kubo:1957mj}. 
Specifically, the retarded Green's function, which describes the causal response of an observable $A$ to a perturbation that couples to an observable $B$, is given by
\begin{equation}
    G_R(t_1,t_2) = -i\theta(t_1-t_2) 
    \langle [A(t_1), B(t_2)] \rangle,
\end{equation}
where $\theta(t_1-t_2)$ is the step function, and the expectation value is taken in the state of interest. The unequal-time commutators computed below therefore serve as the building blocks for the response theory of the model. 

According to Eqs.~(\ref{x0}), (\ref{y0}) and the commutation relation Eq.~(\ref{a_com}), we calculate
\begin{equation}\label{com_xx_0}
    [x_0(t_1), x_0(t_2)] = \tfrac{1}{2} 
    ( [a,a^\da] e^{-i\o(t_1-t_2)} 
    + [a^\da,a]e^{i\o(t_1-t_2)}  )
    = -i \sin [\o(t_1-t_2)],
\end{equation}
\begin{equation}\label{com_xy_0}
    [x_0(t_1), y_0(t_2)] = \tfrac{-i}{2} 
    ( -[a,a^\da] e^{-i\o(t_1-t_2)} 
    + [a^\da,a]e^{i\o(t_1-t_2)}  )
    = i \cos [\o(t_1-t_2)],
\end{equation}
and
\begin{equation}\label{com_yy_0}
    [y_0(t_1), y_0(t_2)] = \tfrac{-1}{2} 
    ( -[a,a^\da] e^{-i\o(t_1-t_2)} 
    -[a^\da,a]e^{i\o(t_1-t_2)}  )
    = -i \sin [\o(t_1-t_2)].
\end{equation}

\subsection{$\l > 0$}

\subsubsection{Canonical momentum and Hamiltonian}

If $\l \ne 0$, constraints are absent. And
we solve $\dot{x}$ and $\dot{y}$ in terms of $x$, $y$, $\pi_x$ and $\pi_y$ from Eqs.~(\ref{canonical momentum x}) and (\ref{canonical momentum y})
\begin{align}
    \dot{x} &= \tfrac{1}{\l} (\pi_x - \tfrac{1}{2} y), \\
    \dot{y} &= \tfrac{1}{\l} (\pi_y + \tfrac{1}{2} x).
\end{align}
The Hamiltonian of this system is
\begin{equation}\label{Hamiltonian}
    H 
    = \dot{x} \pi_x + \dot{y} \pi_y - \mc{L}
    = \frac{\o}{2} (x^2 + y^2) 
    + \frac{1}{2\l}
    \bigl[ (\pi_x - \tfrac{1}{2}y)^2 
    + (\pi_y + \tfrac{1}{2}x)^2 \bigr].
\end{equation}

\subsubsection{Exact solution}

The EOMs determined by the Lagrangian are
\begin{align}
     - \dot{y} - \o x - \l \ddot{x} &= 0, \label{Eom1}  \\
     \dot{x} - \o y - \l \ddot{y} &= 0. \label{Eom2}
\end{align}
We eliminate one variable from above equations to arrive at
\begin{equation}
    \l^2 \ddddot{x} + (1 + 2 \l \o)\ddot{x} + \o^2 x = 0.
\end{equation}
We come up with the trial solution $x(t) = \exp(ift)$, and obtain the characteristic equation
\begin{equation}
    \l^2 f^4 - (1+ 2 \l \o) f^2 + \o^2 = 0.
\end{equation}
We then solve two characteristic frequencies $f_{1}$ and $f_{2}$ of the system
\begin{align}
    f_{1} 
    & = \frac{\sqrt{1+4\l\o}-1}{2\l}, \label{f_1} \\
    f_{2} 
    & =  \frac{\sqrt{1+4\l\o}+1}{2\l}. \label{f_2}
\end{align}
And the general solution to EOMs is given by
\begin{equation}
    \begin{aligned}\label{exact_solution}
        x(t) &= 
        \frac{1}{\sqrt{2\sqrt{1+4\l\o}}}  
        (a_1 e^{-i f_1 t} + a_1^{\da} e^{i f_1 t})
        + \frac{1}{\sqrt{2\sqrt{1+4\l\o}}}  
        (a_2 e^{-i f_2 t} + a_2^{\da} e^{if_2 t}),  \\
        y(t) &= 
        \frac{i}{\sqrt{2\sqrt{1+4\l\o} }} 
        (a_1 e^{-i f_1 t} - a_1^{\da} e^{if_1 t}) 
        + \frac{-i}{\sqrt{2\sqrt{1+4\l\o}}} 
        (a_2 e^{-i f_2 t} - a_2^{\da} e^{if_2 t}).
    \end{aligned}
\end{equation}
Where $a_1, a_1^\da, a_2$ and $a_2^\da$ are complex constants, and the coefficient $\frac{1}{\sqrt{2\sqrt{1+4\l\o}}}$ is imposed for later convenience. 
We also obtain
\begin{align}
    \pi_x(t) &= \tfrac{1}{2}y + \l \dot{x} \nonumber \\
    &= \frac{-i}{2} \sqrt{\frac{\sqrt{1+4\l\o}}{2}}
    (a_1 e^{-i f_1 t} - a_1^{\da} e^{if_1 t}) 
    + \frac{-i}{2} \sqrt{\frac{\sqrt{1+4\l\o}}{2}}
    (a_2 e^{-i f_2 t} - a_2^{\da} e^{if_2 t}),
    \label{pi_x}  \\
    \pi_y(t) &= -\tfrac{1}{2}x + \l \dot{y} \nonumber \\
    &= \frac{1}{2} \sqrt{\frac{\sqrt{1+4\l\o}}{2}}
    (a_1 e^{-i f_1 t} + a_1^{\da} e^{if_1 t})
    - \frac{1}{2} \sqrt{\frac{\sqrt{1+4\l\o}}{2}}
    (a_2 e^{-i f_2 t} + a_2^{\da} e^{if_2 t}).
    \label{pi_y}
\end{align}

\subsubsection{Quantization}
\label{Sec3.2.3}

In order to quantize the system, we promote $x$, $y$, $\pi_x$ and $\pi_y$ to operators, and impose the following equal-time canonical commutation relations
\begin{equation}\label{exact_quantization}
    \begin{aligned}
        \relax [x, \pi_x] &= [y, \pi_y] = i,  \\       
        [x,\pi_y] =[y, \pi_x] 
        &= [x,y] = [\pi_x, \pi_y] 
        = 0.
    \end{aligned}
\end{equation}
Above commutation relations can be achieved as long as the following commutation relations between operators $a_1$, $a_1^\da$, $a_2$ and $ a_2^\da$ are satisfied
\begin{equation}\label{Com_a1a2}
    \begin{aligned}
    \relax [a_1, a_1^\da] &= [a_2, a_2^\da] = 1,  \\
    [a_1, a_2] = [a_1, a_2^\da] 
    &=[a_1^\da, a_2]= [a_1^\da, a_2^\da] = 0.
\end{aligned}
\end{equation}
In terms of operators $a_1, a_1^\da, a_2$ and $a_2^\da$, the Hamiltonian Eq.~(\ref{Hamiltonian}) can be written as
\begin{equation}\label{Hamiltonian_a1a2}
    H = f_1 (a_1^\da a_1 + \tfrac{1}{2}) 
    + f_2 (a_2^\da a_2 + \tfrac{1}{2}).
\end{equation}
The explicit calculation from Eq.~(\ref{exact_quantization}) to Eq.~(\ref{Hamiltonian_a1a2}) is a straightforward but somewhat lengthy exercise. Detailed steps are listed in Appendix~\ref{appB} for completeness.
We define number operators $N_1 \equiv a_1^\da a_1$ and $N_2 \equiv a_2^\da a_2$. Since $N_1$ commutes with $N_2$, we label the common eigenstate of $N_1$ and $N_2$ by $\left| n_1, n_2 \right\rangle$, where $n_1$ and $n_2$ are non-negative integers, i.e.,
\begin{align}
    N_1 | n_1, n_2 \rangle
    &= n_1 | n_1, n_2 \rangle, \\
    N_2 | n_1, n_2 \rangle
    &= n_2 | n_1, n_2 \rangle.
\end{align}
And the energy spectrum of the system is
\begin{equation}
    H \left| n_1, n_2 \right\rangle
    = \big[
    f_1 (n_1 + \tfrac{1}{2})
    + f_2 (n_2 + \tfrac{1}{2})
    \big]
    \left| n_1, n_2 \right\rangle.
\end{equation}

\subsubsection{Unequal-time commutators}

According to the exact solution Eq.~(\ref{exact_solution}) and commutation relations Eq.~(\ref{Com_a1a2}), we calculate
\begin{equation}
    \begin{split}
        [x(t_1), x(t_2)] 
        &= \frac{1}{2\sqrt{1+4\l\o}} 
        \big([a_1, a_1^{\da}] e^{-if_1 (t_1 - t_2)} + [a_1^{\da}, a_1] e^{if_1 (t_1 - t_2)} \big)  \\
        &\quad
        + \frac{1}{2\sqrt{1+4\l\o}} 
        \big([a_2, a_2^{\da}] e^{-if_2 (t_1 - t_2)} + [a_2^{\da}, a_2] e^{if_2 (t_1 - t_2)} \big)  \\
        &=  \frac{-i}{\sqrt{1+4\l\o}} 
        \sin[f_1(t_1 - t_2)]
        + \frac{-i}{\sqrt{1+4\l\o}} 
        \sin[f_2(t_1 - t_2)],
    \end{split}
\end{equation}
\begin{equation}
    \begin{split}
        [x(t_1), y(t_2)] 
        &= \frac{i}{2\sqrt{1+4\l\o}}
        \big(-[a_1, a_1^\da] e^{-if_1 (t_1-t_2)} 
        + [a_1^\da, a_1] e^{if_1 (t_1-t_2)}\big)  \\
        &\quad
        -\frac{i}{2\sqrt{1+4\l\o}} 
        \big(-[a_2, a_2^\da] e^{-if_2 (t_1-t_2)} 
        + [a_2^\da, a_2] e^{if_2 (t_1-t_2)}\big) \\
        &= \frac{-i}{\sqrt{1+4\l\o}} 
        \cos [f_1 (t_1-t_2) ]
        + \frac{i}{\sqrt{1+4\l\o}} 
        \cos [f_2 (t_1-t_2) ],
    \end{split}
\end{equation}
and
\begin{equation}
    \begin{split}
        [y(t_1), y(t_2)] 
        &= \frac{-1}{2\sqrt{1+4\l\o}} 
        \big(-[a_1, a_1^{\da}] e^{-if_1 (t_1 - t_2)} 
        - [a_1^{\da}, a_1] e^{if_1 (t_1 - t_2)} \big) \\
        &\quad
        + \frac{-1}{2\sqrt{1+4\l\o}}
        \big(-[a_2, a_2^{\da}] e^{-if_2 (t_1 - t_2)} 
        - [a_2^{\da}, a_2] e^{if_2 (t_1 - t_2)} \big) \\
        &= 
        \frac{-i}{\sqrt{1+4\l\o}} 
        \sin[f_1(t_1 - t_2)]
        + \frac{-i}{\sqrt{1+4\l\o}} 
        \sin[f_2(t_1 - t_2)].
    \end{split}
\end{equation}

\subsubsection{Behavior as $\l \to 0$}

In the limit $\l \to 0$, $f_1 \to \omega$ and $f_2 \sim 1/\lambda$ diverges. The fast mode decouples from the low-energy physics in two complementary ways. Classically, its frequency diverges, and its contribution to any observable averaged over a macroscopic time interval vanishes. Quantum mechanically, exciting this mode costs an
energy of order $1/\lambda$, which lies far beyond the scope of any low-energy observer. In both regimes, the low-energy physics is governed entirely by the slow mode $f_1$.

To be concrete, we measure the model with accuracy up to $\mc{O}(\l^2)$, 
then $f_1 = \o (1 - \l\o + 2 \l^2 \o^2) + \mc{O}(\l^3)$.
The solution up to $\mc{O}(\l^2)$ is
\begin{equation}\label{xy_behavior}
    \begin{aligned}
        x(t) &= 
        \frac{1-\l\o+\frac{5}{2}\l^2\o^2}{\sqrt{2}} 
        (a_1 e^{-i \o(1-\l\o+2\l^2\o^2) t} + a_1^{\da} e^{i \o(1-\l\o+2\l^2\o^2) t}) + \mc{O}(\l^3),  \\
        y(t) &= 
        -i \frac{1-\l\o+\frac{5}{2}\l^2\o^2}{\sqrt{2}}
        (a_1 e^{-i \o(1-\l\o+2\l^2\o^2) t} - a_1^{\da} e^{i \o(1-\l\o+2\l^2\o^2) t}) + \mc{O}(\l^3).
    \end{aligned}
\end{equation}
Ignoring the high-energy mode, the Hamiltonian up to $\mc{O}(\l^2)$ reads
\begin{equation}
    H = \o (1 - \l \o + 2 \l^2 \o^2) 
    (a_1^\da a_1 + \tfrac{1}{2}) 
    + \mc{O}(\l^3),
\end{equation}
thus the energy spectrum is
\begin{equation}\label{spectrum_behavior}
    H  \left| n_1 \right\rangle
    = \o (1 - \l \o + 2 \l^2 \o^2) 
    (n_1 + \tfrac{1}{2}) 
    \left| n_1 \right\rangle 
    + \mc{O} (\l^3).
\end{equation}
We also obtain unequal-time commutators up to $\mc{O}(\l^2)$
\footnote{Any realistic observation has a finite time resolution $\Delta t$. The measured retarded Green's function is actually a time-averaged version of the exact one. Consequently, the fast mode with frequency much larger than $1 / \Delta t$ is smeared out over a macroscopic time interval. }:
\begin{equation}\label{Com_xx_behavior}
        [x(t_1), x(t_2)] 
        = -i (1 - 2 \l \o + 6 \l^2 \o^2)
        \sin[\o (1 - \l\o + 2\l^2\o^2)(t_1 - t_2)] 
        + \mc{O}(\l^3),
\end{equation}
\begin{equation}\label{Com_xy_behavior}
    [x(t_1), y(t_2)] = -i (1-2\l\o+6\l^2\o^2)
    \cos [\o (1-\l\o+2\l^2\o^2) (t_1-t_2)]
    + \mc{O}(\l^3),
\end{equation}
and
\begin{equation}\label{Com_yy_behavior}
    [y(t_1), y(t_2)] 
    = -i(1 - 2\l \o + 6 \l^2 \o^2) 
    \sin[\o (1-\l\o+2\l^2\o^2)(t_1 - t_2)] 
    + \mc{O}(\l^3).
\end{equation}
The behavior of the exact theory smoothly approaches the case $\l = 0$ as $\l \to 0$. Furthermore, the expansion of the exact theory provides a rigorous benchmark against which the perturbative results obtained in the next section should be tested.

\subsubsection{Difficulties for the perturbative treatment}

The exact theory reduces to the $\l = 0$ case in the limit $\l \to 0$. Thus one would naturally hope for a perturbative quantization scheme that starts from the unperturbed theory and treats the $\l$-term as a perturbation. However, a perturbative treatment of the $\l$-term is not feasible within the standard approach to canonical quantization. The reasons are as follows:

Firstly, the $\l$-term alters the constraint structure of the system. In the unperturbed theory ($\l = 0$), there exist second-class constraints Eqs.~(\ref{constraint1}) and (\ref{constraint2}), and the system lives on a reduced two-dimensional phase space. Quantization is performed by promoting the Dirac brackets, rather than the Poisson brackets, to commutators. In the full theory ($\l > 0$), by contrast, constraints are absent, and the phase space is four-dimensional. Quantization is performed via the canonical commutation relations 
Eq.~(\ref{exact_quantization}). These two quantum theories are structurally distinct---their basic commutation relations are different.

Secondly, the difficulty can also be anticipated from the analytic structure of the Hamiltonian. After the Legendre transform, the Hamiltonian 
Eq.~(\ref{Hamiltonian}) contains a term proportional to $1/\l$, which renders $\l = 0$ a pole of the off-shell Hamiltonian. Therefore it is impossible to split the Hamiltonian into a free part and an interaction part.

In summary, the standard approach to canonical quantization does not provide a scheme in which the $\l$-term can be treated as a perturbation. The covariant phase space formalism, as we will show in the next section, provides precisely such a scheme.

%%%%%%%%%%%%%%%%%%%%%%%%%%%%%%%%%%%%%%%%%%%%%%%%%%%%%%

\section{Perturbative quantization with covariant phase space formalism}\label{Sec4}

In this section we reformulate the quantization of this system with CPS, and treat the $\l$-term as a perturbation. We suppose that we can solve EOMs when $\l = 0$, but the exact solution to the full theory Eq.~(\ref{exact_solution}) were unknown. This is the generic situation in realistic theories, where exact solvability is an exception rather than the rule. Therefore, we solve EOMs by expanding them order by order in $\l$, and quantize this model with CPS based on perturbative solutions\cite{Basu:2004qy}. We reproduce the expansion of the exact theory obtained in the previous section, which provides a non-trivial check for the perturbative method developed with CPS.

\subsection{Symplectic form and Hamiltonian}

The variation of the Lagrangian form $ L =\mc{L} \d t$ is
\begin{equation}
    \de L = 
    (-\dot{y} - \o x - \l \ddot{x}) \de x \d t
    +(\dot{x} - \o y - \l \ddot{y}) \de y \d t 
    + \d \Theta,
\end{equation}
where the symplectic potential $\Theta$ is
\begin{equation}
    \Theta = 
    -\tfrac{1}{2}(x \delta y - y \delta x) 
    + \l (\dot{x} \delta x + \dot{y} \delta y).
\end{equation}
From this we calculate the symplectic form
\begin{equation}\label{Symplectic form}
    \O = \de \Theta
    = -\de x \wedge \de y 
    + \l (\de \dot{x} \w \de x + \de \dot{y} \w \de y).
\end{equation}
The Hamiltonian $H$, which is defined as the Noether current $J_\xi$ associated with the time evolution $\xi = \frac{\del}{\del t}$, is given by\cite{Iyer:1994ys,Harlow:2019yfa}
\begin{equation}\label{Ham_CPS}
    H = J_\xi = X_\xi \cdot \Theta
    - \xi \cdot L
    = \frac{\o}{2}(x^2+y^2) 
    + \frac{\l}{2} (\dot{x}^2 + \dot{y}^2),
\end{equation}
where $X_\xi \equiv \mc{L}_\xi x \frac{\de}{\de x} + \mc{L}_\xi y \frac{\de}{\de y}$ is a vector field defined on the configuration space, and $\cdot$ denotes inserting a vector into the first argument of a differential form.

\subsection{Perturbative solution}

Next we solve the EOMs order by order in $\l$. Suppose that the solution to EOMs can be written as the following perturbative expansion
\begin{equation}\label{expansion}
    \begin{aligned}
        x(t) &= x_0(t) + \l x_1(t) + \l^2 x_2(t) + \cdots,  \\
        y(t) &= y_0(t) + \l y_1(t) + \l^2 y_2(t) + \cdots,
    \end{aligned}
\end{equation}
where 
\begin{equation}\label{unperturbed solution}
    \begin{aligned}
        x_0(t) &= \tfrac{1}{\sqrt2} 
        ( a e^{-i \o t} + a^{\da} e^{i \o t} ),
          \\
        y_0(t) &= \tfrac{-i}{\sqrt2} 
        ( a e^{-i \o t} - a^{\da} e^{i \o t} )
    \end{aligned}
\end{equation}
is the solution to the unperturbed theory, and $x_1(t)$, $y_1(t) $, $x_2(t)$ and $y_2(t)$ are corrections remain to be solved in perturbation theory. Substituting Eq.~(\ref{expansion}) into Eqs.~(\ref{Eom1}) and (\ref{Eom2}), to first order in $\l$ we obtain
\begin{align}
    - \dot{y}_1 - \o x_1 
    &= \ddot{x}_0(t) = - \o^2 x_0(t), \\
    \dot{x}_1 - \o y_1 
    &= \ddot{y}_0(t) = - \o^2 y_0(t).
\end{align}
One specific solution to above equations is
\begin{equation}\label{specific solution 1}
    \begin{aligned}
        x_1(t) &= - \o x_0(t) - \o^2 t y_0(t),  \\
        y_1(t) &= - \o y_0(t) +  \o^2 t x_0(t).
    \end{aligned}
\end{equation}
Thus the general solution to Eqs.~(\ref{Eom1}) and (\ref{Eom2}), valid to first order in $\l$, is
\begin{equation}
    \begin{aligned}
        x(t) &= (1-\l\o) x_0(t) - \l \o^2 t y_0(t) 
        + \mc{O}(\l^2),  \\
        y(t) &= (1-\l\o) y_0(t) + \l \o^2 t x_0(t) 
        + \mc{O}(\l^2).
    \end{aligned}
\end{equation}
We then keep track of the second-order expansions of Eqs.~(\ref{Eom1}) and (\ref{Eom2}), and obtain
\begin{align}
    - \dot{y}_2 - \o x_2 
    &= \ddot{x}_1(t) 
    = 3 \o^3 x_0(t) + \o^4 t y_0(t),  \\
    \dot{x}_2 - \o y_2 
    &= \ddot{y}_1(t) 
    = 3 \o^3 y_0(t) - \o^4 t x_0(t).
\end{align}
One specific solution is
\begin{equation}\label{specific solution 2}
    \begin{aligned}
        x_2(t) &= \tfrac{5}{2} \o^2 x_0(t) 
        + 3 \o^3 t y_0(t) 
        - \tfrac{1}{2} \o^4 t^2 x_0(t), \\
        y_2(t) &= \tfrac{5}{2} \o^2 y_0(t) 
        - 3 \o^3 t x_0(t) 
        - \tfrac{1}{2} \o^4 t^2 y_0(t).
    \end{aligned}
\end{equation}
Thus the general solution to Eqs.~(\ref{Eom1}) and (\ref{Eom2}) valid to second order in $\l$ is
\footnote{We emphasize that this perturbative solution is only valid up to a finite time duration $|t| \ll \frac{1}{\l \o^2}$.}
\begin{equation}\label{perturbative_solution}
    \begin{aligned}
        x(t) &= (1 - \l \o + \tfrac{5}{2} \l^2 \o^2) x_0(t)
        - \l \o^2 t y_0(t)
        + \l^2 (3 \o^3 t y_0(t) 
        - \tfrac{1}{2} \o^4 t^2 x_0(t))
        + \mc{O}(\l^3),  \\
        y(t) &= (1 - \l \o + \tfrac{5}{2} \l^2 \o^2) y_0(t)
        + \l \o^2 t x_0(t)
        + \l^2 (-3 \o^3 t x_0(t) - \tfrac{1}{2} \o^4 t^2 y_0(t))
        + \mc{O}(\l^3).
    \end{aligned}
\end{equation}
We explain that the specific solutions Eqs.~(\ref{specific solution 1}) and (\ref{specific solution 2}) are determined so that the operators $a$ and $a^\da$ satisfy the standard creation-annihilation algebra upon quantization. In the finite time duration 
$|t| \ll \frac{1}{\l \o^2}$ 
the perturbative solution is equivalent to Eq.~(\ref{xy_behavior}), which is obtained by expanding the exact solution in $\l$. In principle we can meet the accuracy we expect by solving EOMs order by order with the perturbative method developed above.

Before proceeding to quantization, we address a natural concern. When $\l > 0$, the perturbed EOMs Eqs.~(\ref{Eom1}) and (\ref{Eom2}) are of second order in time, whereas the unperturbed EOMs Eqs.~(\ref{unperturbed_EOM1}) and (\ref{unperturbed_EOM2}) are of first order. The perturbation therefore signals the presence of an additional degree of freedom---a fast mode whose frequency diverges as $1/\l$ in the full theory. Our perturbative expansion Eq.~(\ref{expansion}), which starts from the $\l = 0$ solution and solves the EOMs order by order in $\l$, generates only the slow mode that is continuously connected to the unperturbed dynamics. The fast mode is invisible in the perturbative expansion because its frequency is not analytic in $\l$ at $\l = 0$. Therefore it is a non-perturbative effect, and it cannot be captured by any finite-order polynomial in $\lambda$. For the purposes of this work, this is precisely what is needed: the fast mode decouples from the low-energy physics in the limit $\l \to 0$, and the perturbative framework faithfully delivers the behavior of the observable slow mode without any prior knowledge of the exact solution.

\subsection{Quantization}\label{Perturbative Quantization}

Substituting the second-order perturbative solution Eq. (\ref{perturbative_solution}) into the symplectic form, all $\mathcal{O}(\l)$ and $\mathcal{O}(\l^2)$ corrections cancel identically. The detailed calculations are given in Appendix~\ref{appC}, and the result is
\begin{equation}\label{symplectic_form_aa_dag}
    \O = -\de x_0 \w \de y_0 + \mc{O}(\l^3)
    = -i \de a \w \de a^\da + \mc{O}(\l^3).
\end{equation}
In order to quantize the system, we promote $a$ and $a^\da$ to operators, and impose the following commutation relation
\begin{equation}\label{CPS_quantization}
    [a,a^\da] =i\O^{-1}(\de a, \de a^\da)= 1.
\end{equation}
We then substitute the perturbative solution Eq. (\ref{perturbative_solution}) into the Hamiltonian Eq. (\ref{Ham_CPS}). After a straightforward simplification (detailed in Appendix~\ref{appC}), we obtain
\begin{equation}\label{perturbed_Hamiltonian}
    H = \omega (1 - \l \o + 2\l^2 \o^2) 
    (a^\da a + \tfrac{1}{2}) + \mc{O}(\l^3).
\end{equation}
We define the number operator $N \equiv a^\da a$ and label its eigenstate by $\left| n \right\rangle$
\begin{equation}
    N | n \rangle
    = n | n \rangle,
\end{equation}
where $n$ is a non-negative integer.
Thus the energy spectrum of the perturbed system is
\begin{equation}
    H | n \rangle
    = \o (1- \l \o + 2 \l^2 \o^2) (n + \tfrac{1}{2})
    | n \rangle
    + \mc{O}(\l^3).
\end{equation}
And this is equivalent to the expansion of the exact theory Eq.~(\ref{spectrum_behavior}).

\subsection{Unequal-time commutators}

Next we compute the unequal-time commutators
within the perturbative framework developed above.
Based on the perturbative solution Eq.~(\ref{perturbative_solution}) and the commutation relation between operators $a$ and $a^\da$ Eq.~(\ref{CPS_quantization}), we calculate
\begin{equation}
    \begin{split}
        [x(t_1), x(t_2)] &= 
        (1 - 2\l\o + 6\l^2\o^2) [x_0 (t_1), x_0 (t_2)]  \\
        &\quad 
        -(1 - 4\l\o) \l\o^2 t_2 [x_0(t_1), y_0(t_2)] 
        -(1 - 4\l\o) \l\o^2 t_1 [y_0(t_1), x_0(t_2)]  \\
        &\quad 
        - \tfrac{1}{2} \l^2 \o^4 t_2^2 
        [x_0 (t_1), x_0 (t_2)] 
        - \tfrac{1}{2} \l^2 \o^4 t_1^2
        [x_0 (t_1), x_0 (t_2)]  \\
        &\quad+ \l^2 \o^4 t_1 t_2 [y_0(t_1), y_0(t_2)] 
        + \mc{O}(\l^3)
        \\
        &= -i (1 - 2\l\o + 6\l^2\o^2) \sin [\o (t_1 - t_2)]
        + i \l\o^2 (1 - 4 \l\o) (t_1 - t_2) 
        \cos [\o (t_1 - t_2)] \\
        &\quad 
        + \tfrac{i}{2} \l^2 \o^4 (t_1 - t_2)^2
        \sin[\o(t_1 - t_2)]
        +\mathcal{O}(\l^3)  \\
        &= -i (1 - 2 \l\o + 6 \l^2\o^2) 
        \sin [\o (1 - \l\o + 2\l^2\o^2) (t_1 - t_2)] +\mathcal{O}(\l^3),
    \end{split}
\end{equation}
\begin{equation}
    \begin{split}
        [x(t_1), y(t_2)] 
        & = (1 - 2\l\o + 6\l^2\o^2) [x_0(t_1), y_0(t_2)] \\
        &\quad 
        + \l\o^2 (1 - 4\l\o) t_2 [x_0(t_1), x_0(t_2)] 
        - \l \o^2 (1 - 4\l\o) t_1 [y_0(t_1), y_0(t_2)]  \\
        &\quad 
        - \tfrac{1}{2} \l^2 \o^4 t_1^2  
        [x_0 (t_1), y_0 (t_2)] 
        - \tfrac{1}{2} \l^2 \o^4 t_2^2 
         [x_0 (t_1), y_0 (t_2)]   \\
        &\quad - \l^2\o^4 t_1 t_2 [y_0(t_1), x_0(t_2)]
        +\mc{O}(\l^3)  \\
        &= i (1 - 2\l\o + 6\l^2\o^2) 
        \cos[\o (t_1 - t_2)] 
        + i\l\o^2 (1 - 4\l\o) (t_1 - t_2) 
        \sin [\o (t_1 - t_2)]   \\
        &\quad 
        - \tfrac{i}{2} \l^2 \o^4 (t_1 - t_2)^2
        \cos [\o(t_1 - t_2)]
        +\mathcal{O}(\l^3) \\
        &= i (1 - 2\l\o + 6\l^2\o^2) 
        \cos[\o (1 - \l\o + 2\l^2\o^2) (t_1 - t_2)] +\mathcal{O}(\l^3),
    \end{split}
\end{equation}
\begin{equation}
    \begin{split}
        [y(t_1), y(t_2)] 
        &=(1 - 2\l\o + 6\l^2\o^2) [y_0 (t_1), y_0 (t_2)] \\
        &\quad 
        + \l\o^2 (1 - 4\l\o) t_2 [y_0(t_1), x_0(t_2)] 
        + \l\o^2 (1 - 4\l\o) t_1 [x_0(t_1), y_0(t_2)]  \\
        &\quad 
        - \tfrac{1}{2} \l^2 \o^4 t_1^2 [y_0 (t_1), y_0 (t_2)]
        - \tfrac{1}{2} \l^2 \o^4 t_2^2 [y_0 (t_1), y_0 (t_2)]\\
        &\quad + \l^2\o^4 t_1 t_2 [x_0(t_1), x_0(t_2)]
        +\mathcal{O}(\l^3) \\
        &= -i (1 - 2\l\o + 6\l^2\o^2) \sin [\o (t_1 - t_2)]
        + i \l\o^2(1 - 4\l\o) (t_1 - t_2) 
        \cos [\o (t_1 - t_2)]   \\
        &\quad 
        +\tfrac{i}{2} \l^2\o^4 (t_1 - t_2)^2 
        \sin [\o (t_1 - t_2)] +\mathcal{O}(\l^3)  \\
        &= -i (1 - 2 \l\o + 6\l^2\o^2) 
        \sin [\o (1 - \l\o + 2\l^2\o^2) (t_1 - t_2)] +\mathcal{O}(\l^3).
    \end{split}
\end{equation}
Where in above calculations we have the expansions of
$\sin[\omega(1-\lambda\omega+2\lambda^2\omega^2)\Delta t]$ 
and 
$\cos[\omega(1-\lambda\omega+2\lambda^2\omega^2)\Delta t]$ 
to second order in $\l$. These results match Eqs.~(\ref{Com_xx_behavior}), (\ref{Com_xy_behavior}) and (\ref{Com_yy_behavior}).

%%%%%%%%%%%%%%%%%%%%%%%%%%%%%%%%%%%%%%%%%%%%%%%%%%%%%%

\section{Conclusion and discussion}\label{Sec5}

In this paper, we have demonstrated how the CPS provides a scheme for the perturbative canonical quantization of an exactly solvable model. This model, although simple, captures the essential difficulties that arise when a velocity-dependent perturbation modifies the kinetic structure of a constraint theory: the $\l \to 0$ limit of the theory is singular, and the perturbed and unperturbed theories live on phase spaces of different dimensions. As a result, treating $\l$-term as a perturbation within the standard approach to canonical quantization is not feasible.

The CPS parries those difficulties by constructing the symplectic form directly on the solution space. The symplectic form in CPS automatically captures the canonical structure of the system, without requiring a prior identification of momenta and constraints. By solving the EOMs order by order in $\l$ and promoting the inverse of the symplectic form to a commutator, we obtained the perturbed quantum theory that agrees with the expansion of the low-energy sector of the exact quantum theory. We verified this agreement explicitly up to second order in $\l$, both for the energy spectrum and for the unequal-time commutators.

Our main methodological conclusion is that the CPS is not merely a conceptual tool for diffeomorphism-invariant theories, but can also serve as a practical and efficient procedure for the perturbative canonical quantization. In the effective field theory framework, perturbations that modify the kinetic structure are understood as low-energy remnants of a more fundamental theory. The perturbative expansion captures only the low-energy degrees of freedom, and the fast modes that lie beyond the cutoff are expected to decouple. The CPS quantization procedure, by constructing the symplectic form directly on this low-energy sector, naturally fits into this physical picture. It is therefore particularly suitable for systems in which perturbations modify the kinetic structure of the unperturbed theory.

Extending this formalism to theories with first-class constraints and gauge symmetries is an important direction for future work\cite{Henneaux:1990au,Henneaux:1992,Pons:2004pp}. We hope that the perturbative quantization of this model will serve as a useful framework for developing perturbative quantization procedures for more complex and physically interesting systems, such as gauge theories, and higher-derivative gravity theories\cite{Cheng:2001du,Cheng:2002rz,Andrzejewski:2007yhi,Dong:2025orj}---including those whose consistency relies on a subtle constraint structure that avoids the Ostrogradsky instability\cite{Woodard:2015zca,Woodard:2006nt,Langlois:2015cwa,Ganz:2020skf}.

%%%%%%%%%%%%%%%%%%%%%%%%%%%%%%%%%%%%%%%%%%%%%%%%%%%%%%

\section*{Acknowledgments}

We thank Xiao-Shuai Wang for helpful discussions. This work is supported by the National Natural Science Foundation of China (NSFC) Project No.12447101 and No.12575079.

%%%%%%%%%%%%%%%%%%%%%%%%%%%%%%%%%%%%%%%%%%%%%%%%%%%%%%

\appendix

\section{Calculations of the Dirac brackets}\label{appA}

In this appendix we calculate the Dirac brackets Eq.~(\ref{Dirac bracket}) between canonical variables. 

The inverse of the $C$ matrix in Section~\ref{Sec3.1.1} is
\begin{equation}
C^{-1} = \begin{pmatrix} 0 & 1 \\ -1 & 0 \end{pmatrix}.
\end{equation}
For the canonical variables $x$, $y$, $\pi_x$ and $\pi_y$, their Poisson brackets with the constraints $\chi_1$ and $\chi_2$ are
\begin{align}
[x, \chi_1]_{\mr{P}} &= 1, & [x, \chi_2]_{\mr{P}} &= 0, \\
[y, \chi_1]_{\mr{P}} &= 0, & [y, \chi_2]_{\mr{P}} &= 1, \\
[\pi_x, \chi_1]_{\mr{P}} &= 0, 
& [\pi_x, \chi_2]_{\mr{P}} & = - \tfrac{1}{2},  \\
[\pi_y, \chi_1]_{\mr{P}} &=  \tfrac{1}{2}, 
& [\pi_y, \chi_2]_{\mr{P}} &= 0.
\end{align}
We then evaluate Dirac brackets of canonical variables
\begin{align}
    [x,y]_{\mr{D}} 
    &= [x,y]_\mr{P} 
    -\sum_{N,M} [x,\chi_N]_\mr{P} 
    (C^{-1})^{NM} 
    [\chi_M, y]_\mr{P}
    = 0 - [1 * 1 * (-1) + 0]
    = 1,  \\
    [x,\pi_x]_{\mr{D}} 
    &= [x,\pi_x]_\mr{P} 
    -\sum_{N,M} [x,\chi_N]_\mr{P} 
    (C^{-1})^{NM} 
    [\chi_M, \pi_x]_\mr{P}
    = 1- [1 * 1 * \tfrac{1}{2} + 0 ]
    =\tfrac{1}{2}, \\
    [x,\pi_y]_{\mr{D}} 
    &= [x,\pi_y]_\mr{P} 
    - \sum_{N,M} [x,\chi_N]_\mr{P} 
    (C^{-1})^{NM} 
    [\chi_M, \pi_y]_\mr{P}
    = 0 - [0 + 0]
    = 0, \\
    [y,\pi_x]_{\mr{D}} 
    &= [y,\pi_x]_\mr{P} 
    - \sum_{N,M} [y,\chi_N]_\mr{P} 
    (C^{-1})^{NM} 
    [\chi_M, \pi_x]_\mr{P}
    = 0 - [0 + 0]
    = 0,  \\
    [y,\pi_y]_{\mr{D}} 
    &= [y,\pi_y]_\mr{P} 
    - \sum_{N,M} [y,\chi_N]_\mr{P} 
    (C^{-1})^{NM} 
    [\chi_M, \pi_y]_\mr{P}
    = 1 - [0 + 1*(-1)*(-\tfrac{1}{2})]
    = \tfrac{1}{2}, \\
    [\pi_x,\pi_y]_{\mr{D}} 
    &= [\pi_x,\pi_y]_\mr{P} 
    - \sum_{N,M} [\pi_x,\chi_N]_\mr{P} 
    (C^{-1})^{NM} 
    [\chi_M, \pi_y]_\mr{P}
    =0 - [(-\tfrac{1}{2})*(-1)*(-\tfrac{1}{2})]
    = \tfrac{1}{4}.
\end{align}

%%%%%%%%%%%%%%%%%%%%%%%%%%%%%%%%%%%%%%%%%%%%%%%%%%%%%%

\section{Details of the exact quantization}\label{appB}

In this appendix calculate the canonical commutation relations and the Hamiltonian given in Section~\ref{Sec3.2.3}.

Firstly we show that commutation relations between operators $a_1$, $a_1^\da$, $a_2$ and $a_2^\da$ given by Eq.~(\ref{Com_a1a2}) imply the canonical commutation relations 
Eq.~(\ref{exact_quantization}). From the exact solution Eq.~(\ref{exact_solution}) and the conjugate momenta Eqs.~(\ref{pi_x}) and (\ref{pi_y}), it is obvious that 
$[x, \pi_y] = [y, \pi_x] = 0$. 
We then calculate 
\begin{align}
        [x, y] &= \frac{i}{2\sqrt{1+4\l\o}} 
        (-[a_1, a_1^\da] + [a_1^\da, a_1])
        - \frac{i}{2\sqrt{1+4\l\o}} 
        (-[a_2, a_2^\da] + [a_2^\da, a_2])
        \nonumber \\
        &= -\frac{i}{\sqrt{1+4\l\o}} 
        + \frac{i}{\sqrt{1+4\l\o}}
        = 0, \\
        [x, \pi_x] &= \tfrac{-i}{4} 
        (-[a_1, a_1^\da] + [a_1^\dagger, a_1])
        + \tfrac{-i}{4} 
        (-[a_2, a_2^\dagger] + [a_2^\dagger, a_2]) 
        = i,  \\
        [y, \pi_y] &= \tfrac{i}{4}
        ([a_1, a_1^\dagger] - [a_1^\dagger, a_1])
        +\tfrac{i}{4} ([a_2, a_2^\dagger] - [a_2^\dagger, a_2]) = i,  \\
        [\pi_x, \pi_y] &= 
        -\tfrac{i}{4} \frac{\sqrt{1+4\l\o}}{2} 
        ([a_1, a_1^\dagger] - [a_1^\dagger, a_1])
        + \tfrac{i}{4} \frac{\sqrt{1+4\l\o}}{2}
        ([a_2, a_2^\dagger] - [a_2^\dagger, a_2]) 
        \nonumber \\
        &= -\tfrac{i}{4} \sqrt{1 + 4 \l \o}
        + \tfrac{i}{4} \sqrt{1 + 4 \l \o}
        =0.
\end{align}

Secondly we show that after quantization, the Hamiltonian Eq.~(\ref{Hamiltonian}) can be re-expressed in terms of operators $a_1$, $a_1^\da$, $a_2$ and $a_2^\da$ as Eq.~(\ref{Hamiltonian_a1a2}). Since the Hamiltonian does not depend on time $t$, for simplicity we substitute the expressions of $x(0)$, $y(0)$, $\pi_x(0)$ and $\pi_y(0)$ into Eq.~(\ref{Hamiltonian}), and obtain
\begin{equation}
\begin{split}
H &= \frac{\o}{4\sqrt{1+4\l\o}} 
     \bigl[
      (a_1 + a_1^\da)^2
    + (a_2 + a_2^\da)^2
    + 2 (a_1 + a_1^\da) (a_2 + a_2^\da) \bigr] \\
    &\quad + \frac{\o}{4\sqrt{1+4\l\o}}  
    \bigl[
    -(a_1 - a_1^\da)^2
    - (a_2 - a_2^\da)^2
    + 2 (a_1 - a_1^\da) (a_2 - a_2^\da) \bigr] 
    \\
  &\quad + \frac{\l}{4\sqrt{1+4\l\o}}
  \bigl[
     -f_1^2 (a_1 - a_1^\da)^2
     -f_2^2 (a_2 - a_2^\da)^2 
     -2 f_1 f_2 (a_1 - a_1^\da) (a_2 - a_2^\da)
     \bigr] \\
  &\quad + \frac{\l}{4\sqrt{1+4\l\o}}
  \bigl[
     f_1^2 (a_1 + a_1^\da)^2
    + f_2^2 (a_2 + a_2^\da)^2
    - 2 f_1 f_2 (a_1 + a_1^\da) (a_2 + a_2^\da)
    \bigr].
\end{split}
\end{equation}
Notice that 
\begin{equation}
    f_1 f_2 = \frac{1 + 4 \l \o - 1}{4\l^2} 
    = \frac{\o}{\l},
\end{equation}
thus terms containing $(a_1 + a_1^\da)(a_2 + a_2^\da)$ and 
$(a_1 - a_1^\da)(a_2 - a_2^\da)$ cancel out in the Hamiltonian. Next we use identities 
$(a+a^\da)^2 = (a^2 + a^{\da 2}) + 2a^\da a + 1$ and
$(a-a^\da)^2 = (a^2 + a^{\da 2}) - 2a^\da a - 1$ to arrive at
\begin{equation}
\begin{split}
    H&= \frac{\o}{2\sqrt{1+4\l\o}} \bigl[
    (2 a_1^\da a_1 + 1)
     + (2 a_2^\da a_2 + 1)
    \bigr] \\
    &\quad + \frac{\l}{2\sqrt{1+4\l\o}} \bigl[
      f_1^2 (2 a_1^\da a_1 + 1)
    + f_2^2 (2 a_2^\da a_2 + 1)
    \bigr] \\[2pt]
    &= \frac{\o + \l f_1^2}{\sqrt{1+4\l\o}} 
    (a_1^\da a_1 + \tfrac{1}{2})
    + \frac{\o + \l f_2^2}{\sqrt{1+4\l\o}} 
    (a_2^\da a_2 + \tfrac{1}{2}). 
\end{split}
\end{equation}
Finally we calculate
\begin{equation}
    \o + \l f_1^2 = \o 
    + \frac{(1 +  2\l\o) - \sqrt{1+ 4\l\o}}{2\l}
    = \frac{(1 +  4\l\o) - \sqrt{1+ 4\l\o}}{2\l}
    = \sqrt{1+ 4\l\o} f_1,
\end{equation}
\begin{equation}
    \o + \l f_2^2 = \o 
    + \frac{(1 +  2\l\o) + \sqrt{1+ 4\l\o}}{2\l}
    = \frac{(1 +  4\l\o) + \sqrt{1+ 4\l\o}}{2\l}
    = \sqrt{1+ 4\l\o} f_2.
\end{equation}
As a consequence, we obtain
\begin{equation}
    H = f_1 (a_1^\da a_1 + \tfrac{1}{2}) 
    + f_2 (a_2^\da a_2 + \tfrac{1}{2}).
\end{equation}

%%%%%%%%%%%%%%%%%%%%%%%%%%%%%%%%%%%%%%%%%%%%%%%%%%%%%%

\section{Details of the perturbative quantization}\label{appC}

In this appendix we calculate the symplectic form and the Hamiltonian in Section~\ref{Perturbative Quantization}. 

Firstly we calculate the symplectic form Eq.~(\ref{symplectic_form_aa_dag}). 
Given the second-order perturbative solution Eq.~(\ref{perturbative_solution}), we calculate
\begin{equation}
    \begin{split}
        \de x \w \de y &= 
        (1 - \l \o + \tfrac{5}{2} \l^2 \o^2)^2 
        \de x_0 \w \de y_0
        - \l^2 \o^4 t^2 \de y_0 \w \de x_0  \\
        &\quad - \tfrac{1}{2} \l^2 \o^4 t^2 \de x_0 \w \de y_0
        - \tfrac{1}{2} \l^2 \o^4 t^2 \de x_0 \w \de y_0 
        + \mc{O}(\l^3)\\
        &= (1 - 2 \l \o + 6 \l^2 \o^2)  \de x_0 \w \de y_0
        + \mc{O}(\l^3),
    \end{split}
\end{equation}
\begin{equation}
    \begin{split}
    \l \de \dot{x} \w \de x
    &= \l \big[(1 - \l \o)\de \dot{x}_0 
    - \l \o^2 \de y_0 
    - \l \o^2 t \de\dot{y}_0\big] \w 
    \big[(1-\l\o) \de x_0 
    - \l \o^2 t \de y_0 \big] + \mc{O}(\l^3) \\
    &= \l \big[ (1 - \l \o) \o \de y_0 
    - \l \o^2 \de y_0
    + \l \o^3 t \de x_0 \big]
    \w \big[(1-\l\o) \de x_0 
    - \l \o^2 t \de y_0 \big] + \mc{O}(\l^3)  \\
    &= \l (1 - \l \o)^2 \o \de y_0 \w \de x_0
    - \l^2 \o^2 \de y_0 \w \de x_0 + \mc{O}(\l^3) \\
    &= - (\l\o - 3 \l^2\o^2) \de x_0 \w \de y_0  + \mc{O}(\l^3),
    \end{split}
\end{equation}
and
\begin{equation}
    \begin{split}
    \l \de \dot{y} \w \de y
    &= \l \big[(1 - \l \o)\de \dot{y}_0 
    + \l \o^2 \de x_0 
    + \l \o^2 t \de\dot{x}_0 \big] \w 
    \big[(1-\l\o) \de y_0 
    + \l \o^2 t \de x_0 \big] + \mc{O}(\l^3) \\
    &= \l \big[-(1 - \l \o) \o \de x_0 
    + \l \o^2 \de x_0
    + \l \o^3 t \de y_0 \big]
    \w \big[(1-\l\o) \de y_0 
    + \l \o^2 t \de x_0 \big] + \mc{O}(\l^3)  \\
    &= -\l (1 - \l \o)^2 \o \de x_0 \w \de y_0
    + \l^2 \o^2 \de x_0 \w \de y_0 + \mc{O}(\l^3) \\
    &= - (\l\o - 3 \l^2\o^2) \de x_0 \w \de y_0  + \mc{O}(\l^3).
    \end{split}
\end{equation}
Where we have used $\de x_0 \w \de x_0 = \de y_0 \w \de y_0 = 0$, and Eqs.~(\ref{unperturbed_EOM1}) and (\ref{unperturbed_EOM2}) in above calculations. Thus we obtain
\begin{equation}
    \begin{split}
        \O &= 
        - \de x \w \de y  
        + \l (\de \dot{x}\w \de x + \de \dot{y}\w \de y)  \\
        &= 
        -(1 - 2 \l \o + 6 \l^2 \o^2) 
        \de x_0 \w \de y_0 
        - 2\l\o \de x_0 \w \de y_0 
        + 6 \l^2 \o^2 \de x_0 \w \de y_0
        + \mathcal{O}(\l^3) \\
        &= - \de x_0 \w \de y_0 + \mathcal{O}(\l^3).
    \end{split}
\end{equation}
We then substitute the unperturbed solution Eq.~(\ref{unperturbed solution}) into the symplectic potential. Finally we obtain
\begin{equation}
\begin{split}
    \O &= - \de x_0 \w \de y_0 + \mc{O}(\l^3)
    = \tfrac{i}{2} (\de a e^{-i \o t} + \de a^\da e^{i \o t})
    \w (\de a e^{-i \o t} - \de a^\da e^{i \o t}) + \mc{O}(\l^3) \\
    &= -i \de a \w \de a^\da + \mc{O}(\l^3).
\end{split}
\end{equation}

Secondly rewrite the Hamiltonian in terms of operators $a$ and $a^\da$ as Eq.~(\ref{perturbed_Hamiltonian}). Since the Hamiltonian does not evolve with time $t$, we evaluate all quantities at $t=0$ for simplicity. We have
\begin{equation}
    x^2(0) = 
    (1 - \l\o + \tfrac{5}{2} \l^2 \o^2)^2 x_0^2(0) +\mc{O}(\l^3)
    = (1 - 2\l\o + 6\l^2\o^2) x_0^2(0)+\mc{O}(\l^3),
\end{equation}
\begin{equation}
    y^2(0) = (1 - \l\o + \tfrac{5}{2} \l^2 \o^2)^2 y_0^2(0) +\mc{O}(\l^3)
    = (1 - 2\l\o + 6\l^2\o^2) y_0^2(0) +\mc{O}(\l^3),
\end{equation}
\begin{equation}
\begin{split}
    \dot{x}^2(0) &= 
    \big[ (1-\l\o) \dot{x}_0(0) - \l \o^2 y_0(0) \big]^2 
    + \mc{O}(\l^2) 
    = (1 - 2 \l \o)^2 \o^2 y_0^2(0) + \mc{O}(\l^2) \\
    &= (1 - 4\l \o) \o^2 y_0^2(0) + \mc{O}(\l^2) ,
\end{split}
\end{equation}
and
\begin{equation}
\begin{split}
    \dot{y}^2(0) &= 
    \big[ (1-\l\o) \dot{y}_0(0) + \l \o^2 x_0(0) \big]^2 
    + \mc{O}(\l^2)  
    = (1 - 2 \l \o)^2 \o^2 x_0^2(0) + \mc{O}(\l^2) \\
    &= (1 - 4\l \o) \o^2 x_0^2(0)+ \mc{O}(\l^2).
\end{split}
\end{equation}
Thus
\begin{equation}
    \begin{split}
        H &= 
        \tfrac{\o}{2} (x^2 + y^2) 
        + \tfrac{\l}{2} (\dot{x}^2 + \dot{y}^2)  \\
        &= 
        \tfrac{\o}{2} 
        (1 - 2\l\o + 6\l^2\o^2) (x_0^2 + y_0^2)
        + \tfrac{\l}{2} 
        (1 - 4\l\o) \o^2 (x_0^2 + y_0^2)  
        + \mc{O}(\l^3)\\
        &= \tfrac{1}{2} \o (1- \l \o + 2 \l^2 \o^2) 
        (x_0^2 + y_0^2) 
        +\mc{O}(\l^3)  \\
        &= \o (1- \l \o + 2 \l^2 \o^2) 
        (a^\da a + \tfrac{1}{2}) 
        +\mc{O}(\l^3).
    \end{split}
\end{equation}

%%%%%%%%%%%%%%%%%%%%%%%%%%%%%%%%%%%%%%%%%%%%%%%%%%%%%%

\addcontentsline{toc}{section}{References}

\bibliographystyle{JHEP}
\bibliography{refs}

\end{document}